\documentclass[a4paper]{jpconf}
\usepackage{graphicx}

\newcommand{\sgra}{Sgr~A*}

\newcommand{\rmm}{rad~m$^{-2}$}
\newcommand{\rin}{$r_\mathrm{in}$}

\newcommand{\msunyr}{$M_\odot$~yr$^{-1}$}

\newcommand{\arcdeg}{$^\circ$}
\newcommand{\arcsec}{\mbox{$^{\prime\prime}$}}
\newcommand\farcs{\mbox{$.\!\!^{\prime\prime}$}}

\begin{document}

\title{The Submillimeter Polarization of Sgr A*}

\author{Daniel P. Marrone$^1$, James M.Moran$^1$, Jun-Hui Zhao$^1$ and
Ramprasad Rao$^2$}

\address{$^1$ Harvard-Smithsonian Center for Astrophysics, 60 Garden
St., Cambridge, MA 02138, USA}
\address{$^2$ Inst. of Ast. and Astrophys., Academia
Sinica, P.O. Box 23-141, Taipei 10617, Taiwan}

\ead{dmarrone@cfa.harvard.edu}

\begin{abstract}
We report on the submillimeter properties of \sgra\ derived from
observations with the Submillimeter Array and its polarimeter. We find
that the spectrum of \sgra\ between 230 and 690~GHz is slightly
decreasing when measured simultaneously, indicating a transition to
optically thin emission around $300-400$~GHz. We also present very
sensitive and well calibrated measurements of the polarization of
\sgra\ at 230 and 345~GHz. With these data we are able to show for the
first time that the polarization of \sgra\ varies on hour timescales,
as has been observed for the total intensity. On one night we find
variability that may arise from a polarized ``blob'' orbiting the
black hole. Finally, we use the ensemble of observations to determine
the rotation measure. This represents the first statistically
significant rotation measure determination and the only one made
without resorting to comparing position angles measured at separate
epochs. We find a rotation measure of $(-5.6\pm0.7)\times10^5$~\rmm,
with no evidence for variability on inter-day timescales at the level
of the measurement error. The stability constrains interday
fluctuations in the accretion rate to 8\%. The mean intrinsic
polarization position angle is 167\arcdeg$\pm$7\arcdeg\ and we detect
variations of $31^{+18}_{-9}$ degrees. This separation of intrinsic
polarization changes and possible rotation measure fluctuations is now
possible because of the frequency coverage and sensitivity of our
data. The observable rotation measure restricts the accretion rate to
the range $2\times10^{-7}$~\msunyr\ to $2\times10^{-9}$~\msunyr, if
the magnetic field is near equipartition and ordered.
\end{abstract}

\section{Introduction}
The linear polarization of \sgra\ was first detected by Aitken et
al. \cite{AitkenE00} above 100~GHz, after unsuccessful searches at
lower frequencies \cite{BowerE99-lp1,BowerE99-lp2}. Through subsequent
observations it has been established that the polarization varies in
position angle \cite{BowerE03,BowerE05} and fraction
\cite{MarroneE06}, with variability occurring on timescales comparable
to those of the previously observed total intensity variations (see
\S~\ref{s-quorbit}). The variability may be intrinsic to the source or
due to propagation effects, but the short timescales suggest processes
at work very close to the black hole and thus polarization should be a
useful tool for the study of \sgra.

The ability to exploit polarization information is significantly
sensitivity-limited for this faint source. Until very recently
polarimetric observations have required full nights to obtain reliable
measurements, obscuring changes on shorter intervals that might be
expected based on total intensity variations. Perhaps more important
has been the limited instantaneous frequency coverage.  Multiple
frequencies must be observed in order to separate the local changes in
the emission region from the external changes in the propagation
medium (such as rotation measure changes). In the absence of spectral
information, these changes can at best be separated statistically
through comparisons of observations from different frequencies and
different times. As a result of the poor frequency coverage available
to date, Faraday rotation has yet to be convincingly detected in this
source. Because of its potential to constrain the accretion rate close
to the black hole, measurement of Faraday rotation has been a chief
goal of polarimetric observations.

Here we present several results obtained with the Submillimeter Array
(SMA). The SMA is a sensitive new instrument at a good submillimeter
site and favorable latitude for observing the Galactic center. Its has
large instantaneous frequency coverage; single-receiver observations
provide data in two sidebands 10~GHz apart and in dual-receiver
observations the frequency span can be a factor of three (230 to
690~GHz). All of these advantages make the SMA a powerful instrument
for the study of the Galactic center, and \sgra\ in particular. We
report on observations of the instantaneous spectrum of
\sgra\ and the peak in its spectral energy distribution
(\S~\ref{s-SED}), the polarization variability on short timescales
(\S~\ref{s-quorbit}), and the first measurement of Faraday rotation in
this source (\S~\ref{s-RM}). Other SMA Galactic center observations
are presented by Montero-Casta\~no, Qin, and Winnberg in this volume.

\section{Submillimeter Spectrum}
\label{s-SED}
\sgra\ is known to have a rising spectrum across three decades in
frequency from 300~MHz to 300~GHz. A few detections have been made at
frequencies up to 900~GHz
\cite{AitkenE00,ZylkaE95,SerabynE97,PiercePriceE00,YZE06}, but the
flux densities are scattered and \sgra\ is occasionally undetectable
at the same telescope and frequency \cite{SerabynE97,DentE93}. This
suggests variability in the source and may also indicate the
difficulty of extracting a flux density for \sgra\ from observations
with $10-20$\arcsec\ angular resolution in the presence of
contaminating dust and free-free emission. There are only upper limits
on the flux density in the far and mid-infrared, while (transient)
counterparts in the near-infrared and X-ray show quiescent flux
densities that are several orders of magnitude below that at 300~GHz
\cite{GenzelE03,GhezE04,BaganoffE03}. These data suggest an unobserved
peak in the spectral energy distribution (SED) between 300~GHz and the
mid-infrared.

\begin{figure}[b]
\includegraphics[width=22pc]{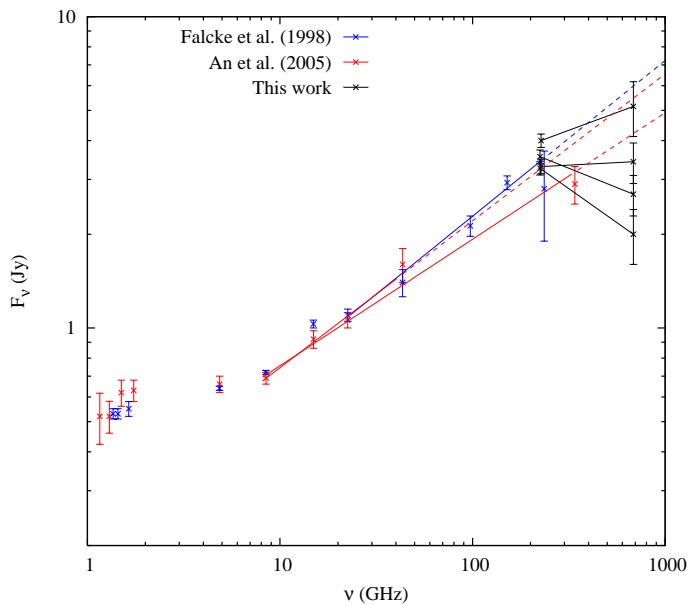}
\hspace{1.5pc}
\begin{minipage}[b]{14pc}
\caption{\label{f-sed}The radio SED of \sgra. Points from two
multi-wavelength campaigns \cite{FalckeE98,AnE05} are shown with
power-law fits to the ``submillimeter bump''. Extrapolated spectra are
continued with dashed lines. We provide two fits to the An et
al. \cite{AnE05} points, with and without the 340~GHz point, as this
frequency is expected to be near the SED peak (and thus below the
extrapolation from lower frequencies) based on our observations. The
230 and 690~GHz points in each epoch of SMA data are connected by a
solid line.}
\end{minipage}
\end{figure}

Using the unprecedented submillimeter angular resolution afforded by
the SMA, as well as its ability to observe simultaneously at 230 and
690~GHz, we have made several observations of the submillimeter
emission of \sgra. The radio spectrum is shown in Figure~\ref{f-sed},
including four epochs of SMA 230/690~GHz observations. The SMA data
were obtained on three nights in 2005, with one night split into two
intervals due to noticeable variability. The data were obtained in
excellent weather, 225~GHz zenith opacities of $0.034-0.045$ and
corresponding 690~GHz opacities of $0.55-0.75$. Through careful
calibration against quasars and multiple primary flux standards
(planets), we were able to determine the flux density scale to 15\%
precision at 690~GHz on the best nights. We find that the spectral
index ($\alpha$, for $\mathrm{S}\propto\nu^\alpha$) between the two
frequencies is flat, ranging between $-0.4$ and $+0.2$, with a
variance-weighted average value of $\alpha=-0.13$.

We contrast these observations to the spectral index measured at lower
frequencies. Below approximately 10~GHz, \sgra\ shows a slowly rising
spectrum \cite{FalckeE98,AnE05,ZhaoBoGo01} typical of an inhomogeneous
optically thick synchrotron source. Closer to our frequencies,
observations have found a steeper spectrum in a ``submillimeter bump''
\cite{MeliaFalcke01} extending to at least 200~GHz. Spectral
measurements at the highest frequencies must be undertaken with care
because interday variations of $20-40$\% are observed above 100~GHz
\cite{ZhaoE03}, with larger variations observed on few hour timescales
\cite{MarroneE06,MiyazakiE04,MauerhanE05}. Two multi-facility
campaigns to measure the simultaneous spectrum of
\sgra\ to $\sim300$~GHz are available to provide calibration for our
results. In the 1996 observations of Falcke et al. \cite{FalckeE98}
the average spectral index from 22 to 236~GHz is 0.50, while the 2003
campaign of An et al. \cite{AnE05} shows an average spectral index of
0.41 from 8.4 to 340~GHz, or 0.47 if the 340~GHz point is
excluded. Yusef-Zadeh et al. \cite{YZE06} also have extensive
wavelength coverage but unfortunately their flux densities are
contaminated by emission from a radio transient \cite{BowerE05-Trans}
and by flaring. The fits to the simultaneous spectra are plotted in
Figure~\ref{f-sed} and demonstrate that most of our 230~GHz points
correspond well to the extrapolated lower-frequency spectra. The
exceptional epoch has a high flux density in both bands and is unusual
for other reasons, as shown in Section~\ref{s-quorbit}. We clearly
observe a deviation from these extrapolations at 690~GHz, with flux
density decrements of up to 4~Jy, indicating that we are straddling
the peak in the SED. In models of the spectrum of this source the
turnover originates from the transition to optically thin emission,
typically in a compact component of synchrotron emission corresponding
to the inner regions of an accretion flow or the base of a jet
\cite{YuanE03,FalckeMarkoff00}. We therefore infer that at frequencies
above approximately 300~GHz we have reached this transition and are
observing the emission closest to the black hole. This result is
particularly important for planned 345~GHz VLBI observations of the
relativistically warped image of the black hole. In the case of
emission from an accretion flow, VLBI images near the transition
frequency should show more structure than would be expected from
observations at very optically thin or optically thick frequencies
\cite{BrodLoeb06-apjl}. Something similar might be expected for a jet
model since the compact jet nozzle should be the dominant emission
source at high frequencies \cite{FalckeMarkoff00}, rather than larger
scale jet emission that would not appear warped by strong gravity.

\section{Polarization Light Curves}
\label{s-quorbit}
The SMA observations are the first to allow short time interval
sampling of the millimeter/submillimeter linear polarization of \sgra\
(see Eckart et al. in this volume for polarimetric data in the near
infrared). Our 2005 data sample the 230~GHz polarization of \sgra\ in
4 minute integration periods, yielding a typical precision of
$0.5-1.0$\%, while the 345~GHz polarization is measured over 5.5
minute periods at a precision of $1.3-1.8$\%. For calibration details
see Marrone \cite{thesis}. This temporal resolution is at least a
factor of thirty better than the only previous time-resolved data
\cite{MarroneE06}. As shown by Broderick and Loeb
\cite{BroderickLoeb06} (see also, these proceedings), the polarization
changes during flares should be an essential signature of strong
gravity. With a favorable flare, our temporal resolution may allow us
to measure the black hole spin and properties of the flaring plasma.

\begin{figure}[t]
\includegraphics[width=24pc]{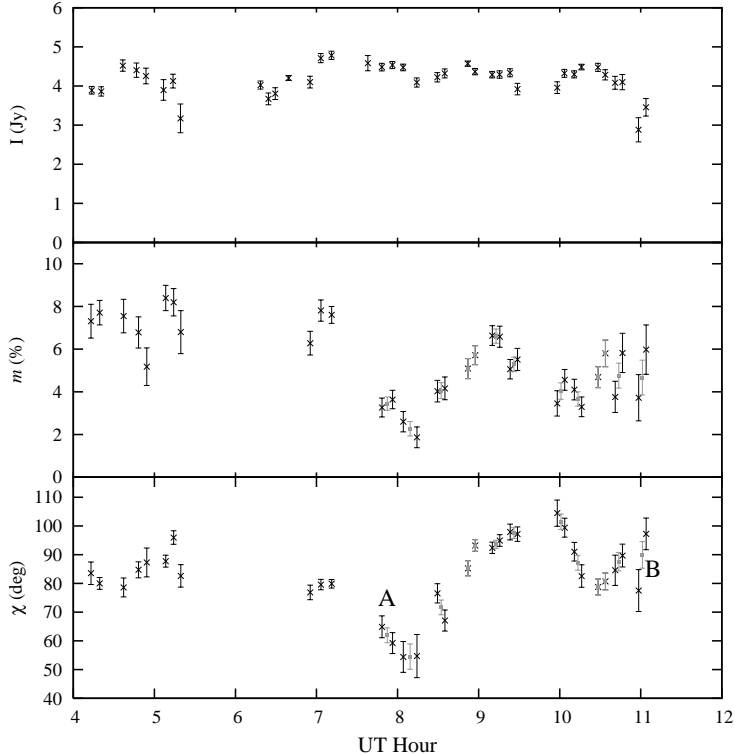}
\hspace{1.5pc}
\begin{minipage}[b]{12pc}
\caption{\label{f-PLC}The variability of \sgra\ in total intensity
(Stokes I), polarization fraction ($m$), and polarization position
angle ($\chi$) over one night of observations. Each black point
corresponds to a four minute sample of the polarization of \sgra. In
the lower two panels the grey points show the same data rebinned to
increase the signal-to-noise ratio; it is these points that are
plotted in Figure~\ref{f-QU}. The grey points labeled A and B in the
lower panel are similarly labeled in Figure~\ref{f-QU}.\vspace{1pc}}
\end{minipage}
\end{figure}

A sample polarized light curve from the SMA is shown in
Figure~\ref{f-PLC}. The data are from the same night as the highest
230/690~GHz points in Figure~\ref{f-sed}. Although the total intensity
(Stokes I) does vary during these observations, the changes visible in
the polarization are much more dramatic. The light curve shows a
50\arcdeg\ swing in polarization position angle over a period of just
two hours and a decrease in polarization fraction from nearly 8\% to
2\% over just one hour. Maintaining relative flux calibration at the
level of a few percent is difficult in the submillimeter, making the
$10-20$\% variability observed in Stokes I difficult to measure to
high significance. However, the polarization information is much
easier to measure reliably because the observations can be
self-calibrated on the relatively strong point source emission from
\sgra\ \cite{thesis}. This makes polarimetry particularly useful for
monitoring observations.

\begin{figure}[t]
\includegraphics[width=24pc]{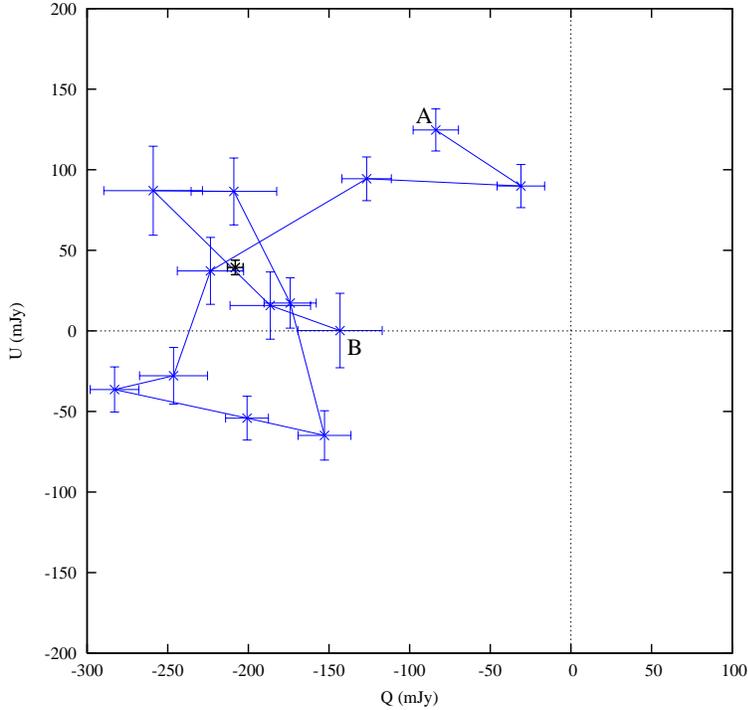}
\hspace{1.5pc}
\begin{minipage}[b]{12pc}
\caption{\label{f-QU}The path traced in the Q$-$U plane over the last
three and a half hours of the observation shown in
Figure~\ref{f-PLC}. Time increases from point A to point B. The mean
polarization for the entire night is shown with a bold cross near
Q$=-210$~mJy, U=40~mJy. The polarization makes nearly two full orbits
around the mean polarization.\vspace{1pc}}
\end{minipage} 
\end{figure}

The polarization variability near the end of the observations in
Figure~\ref{f-PLC} looks vaguely periodic when plotted in $m$ and
$\chi$ coordinates. Examination of more natural coordinates, Stokes Q
and U, confirms that we do observe a polarization orbit. The grey
points from Figure~\ref{f-PLC} are plotted in the Q$-$U plane in
Figure~\ref{f-QU}, with time proceeding from point A to point B. The
polarization appears to circle the mean polarization (marked by the
black point) multiple times over the last hours of the light curve. We
can understand this track with a simple picture of a submillimeter
flare in \sgra. We attribute the flaring emission to a newly excited
blob of synchrotron-emitting plasma orbiting the black hole in the
inner regions of an accretion flow. As the blob orbits the black hole
its apparent linear polarization direction will rotate, so when this
is added vectorially to the polarization of the surrounding quiescent
emission it appears to orbit the mean polarization. The roughly four
hour period (twice the period in the Q$-$U plane) would correspond to
an orbital radius of around 13 Schwarzschild radii ($r_\mathrm{S}$),
well outside the marginally stable orbit for a black hole of any spin
parameter. Although other mechanisms for variability in \sgra\ are
certainly possible, this simple picture matches well with our
observations. More complicated effects such as synchrotron losses,
general relativistic light paths, and radial motion of the blob can
also be considered to allow better fits of the observed polarization
track, as discussed by Broderick and Loeb in this
volume. Nevertheless, the correspondence between this picture and our
observations gives us hope for both the utility of short-timescale
polarimetry in understanding \sgra\ flares and the potential for
modeling polarized flares.

\section{Rotation Measure and Intrinsic Polarization}
\label{s-RM}

Despite great interest, a detection of the Faraday rotation measure
(RM) has remained elusive since the discovery of linear polarization
in 1999 \cite{AitkenE00}. The presence of linear polarization was used
immediately after its detection to argue that the infalling plasma
must be tenuous and the mass accretion rate low (less than
$10^{-6}$~\msunyr), as larger accretion rates would depolarize the
emission through extreme Faraday rotation angles
\cite{QuatGruz00-LP,Agol00}. Further progress on determining the
accretion rate and its variability has awaited an RM measurement,
which is made difficult by the polarization variability and the
diminished effect of Faraday rotation at high frequencies. To date
there have been three claims of a RM determination through
non-simultaneous observations, although none of them is robust. Bower
et al. \cite{BowerE03} used the few available measurements to
determine an RM of $-4.3\times10^5$~\rmm, but the discovery of
polarization variability rendered the result uncertain
\cite{BowerE05}. Marrone et al. \cite{MarroneE06} compared average 230
and 345~GHz position angles to find $-5\times10^5$~\rmm. Most
recently, Macquart et al. \cite{MacquartE06} found an RM of
$-4.4\times10^5$~\rmm\ from their 83~GHz polarization and the average
of all previous data. However, their analysis allows a much lower RM
at a probability of 10\% or more due to the 180\arcdeg\ degeneracy of
the polarization position angle.

Simultaneous measurements are a much more secure way to determine the
RM, but the instantaneous frequency coverage of available instruments
has not been adequate to show Faraday rotation. The lowest
instantaneous RM upper limit is presently $-7\times10^5$~\rmm\ from
early SMA 345~GHz polarimetric observations \cite{MarroneE06},
comparable to the RM expected from non-simultaneous measurements.
Multi-frequency data allow for the determination of both the RM and
intrinsic polarization direction, allowing separation of their
fluctuations. Isolating these changes is crucial to understanding the
conditions in the inner accretion flow. The RM can be used to
determine upper and lower limits on the accretion rate, while
variations with frequency and time can be used to examine the
structure of the flow. In this section we use SMA measurements to
obtain a highly significant RM detection and strong evidence that the
intrinsic polarization is variable.

\subsection{Observations and Calibration}
The results in this section depend crucially on careful calibration
and thorough examination of possible systematic effects. Measurements
of the RM in \sgra\ with the SMA require convincing detections of
inter-sideband position angle differences of 1.5\arcdeg\ at 345~GHz
and 5\arcdeg\ at 230~GHz. Here we present a summary of the
calibration, with details reserved for a paper (Marrone et al., 
submitted to {\it ApJL}).

The data presented here were obtained on ten nights in the summer of
2005 during polarimeter commissioning and as part of regular science
programs. There were six tracks in the 230~GHz band and four in the
345~GHz band. The weather was very good, with a typical zenith opacity
at 225~GHz of $0.06-0.08$, or $0.19-0.27$ when scaled to 343~GHz. All
data were obtained with five to seven antennas in the compact
configuration ($7-70$~meter baselines), resulting in a synthesized
beam width of 2\farcs0$\times$4\farcs0 at 230~GHz and
1\farcs6$\times$3\farcs2 at 345~GHz on \sgra. For the purposes of
self-calibration and polarization extraction, baselines shorter than
$20 k\lambda$ were removed to exclude the extended emission around
\sgra. Polarization was extracted through point-source fits to the
visibilities. To obtain full polarization information we used the SMA
polarimeter \cite{MarroneE06}, which will be described in a
forthcoming instrument paper (D.~P.~Marrone et al., 2006, in
prep.). In 2005 new wave plates were added to the polarimeter to allow
simultaneous polarimetry in the SMA 230~GHz and 690~GHz bands. These
represent the first scientific data obtained with the new
capabilities.

Measurement of polarization relies on precise determination of the
fractional contamination (``leakage'') of each polarization state by
the cross-handed polarization. Uncalibrated leakage contaminates the
linearly polarized Stokes parameters ($Q$ and $U$) with $I$. Leakages
were measured each night by observing polarized quasars (3C~279 or
3C~454.3) over a large range of parallactic angle and simultaneously
solving for the source and instrumental polarization. The leakages are
stable, with an r.m.s. variability of 0.3\% at 230 GHz and 0.4\% at
345~GHz. There are three potential sources of variability: (1) real
instrumental polarization changes between nights, (2) finite
signal-to-noise ratio, and (3) polarization leakage that is not
constant across the sky, resulting in leakage determinations that
depend on the hour angle coverage. The first two sources are likely
present at a low level. The third effect is also present, at a level
of around 0.2\%, due to the SMA optical configuration. Fortunately,
the resulting polarization changes are small. Leakage errors of
0.3\%$-$0.4\% contribute $0.2$\% fractional contamination to the
instantaneous polarization measured with seven antennas, significantly
less when averaged over a full track due to the parallactic angle
rotation. For \sgra, which we measure to be 5\%$-$10\% polarized, this
results in at most 1\arcdeg\ of position angle error. More importantly
for this purpose, the uncorrected instrumental polarization (source 3,
above) is very nearly constant across the sidebands. Therefore,
although the absolute position angle varies by up to 1\arcdeg, the
inter-sideband difference varies only by 0.1\arcdeg$-$0.2\arcdeg.

\subsection{Rotation Measure and Intrinsic Polarization}
\begin{figure}[t]
\begin{minipage}{18pc}
\includegraphics[width=18pc]{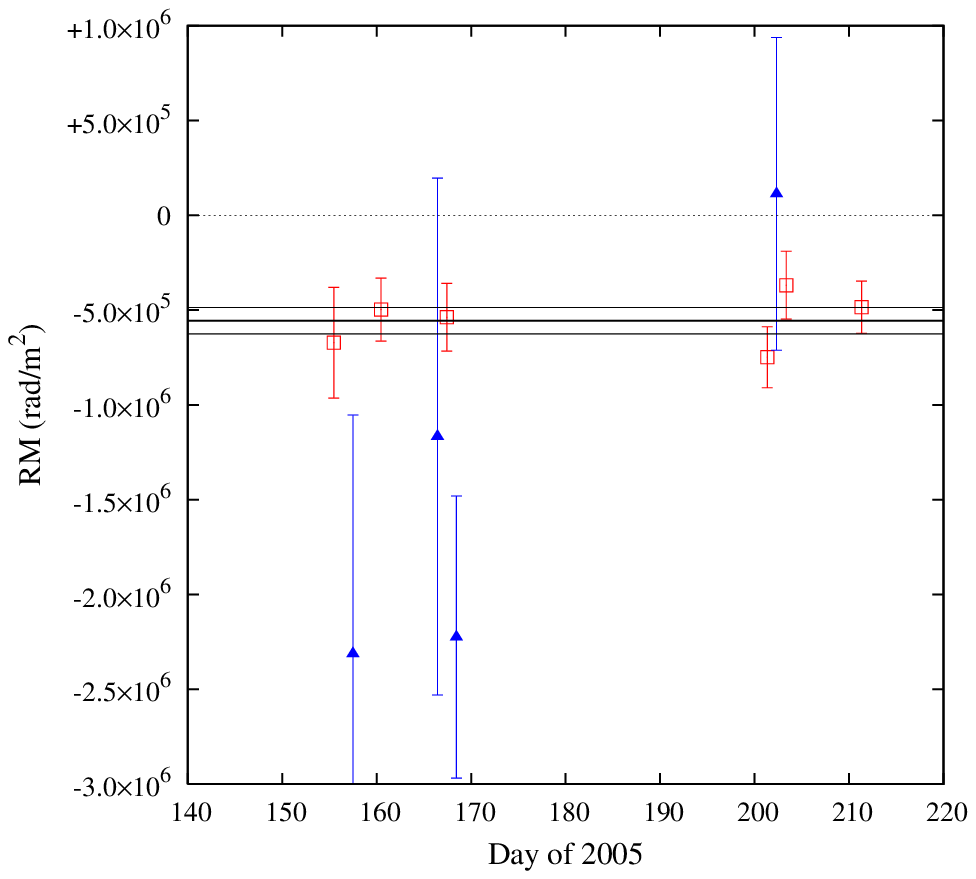}
\caption{\label{f-RM}Rotation measure derived from ten nights of
observations in the summer of 2005. Measurements at 230~GHz are shown
as squares and those at 345~GHz are shown as triangles. The mean RM
($-5.6\times10^5$~\rmm) and $1\sigma$ range are shown with horizontal
lines.}
\end{minipage}
\hspace{1.5pc}
\begin{minipage}{18pc}
\includegraphics[width=18pc]{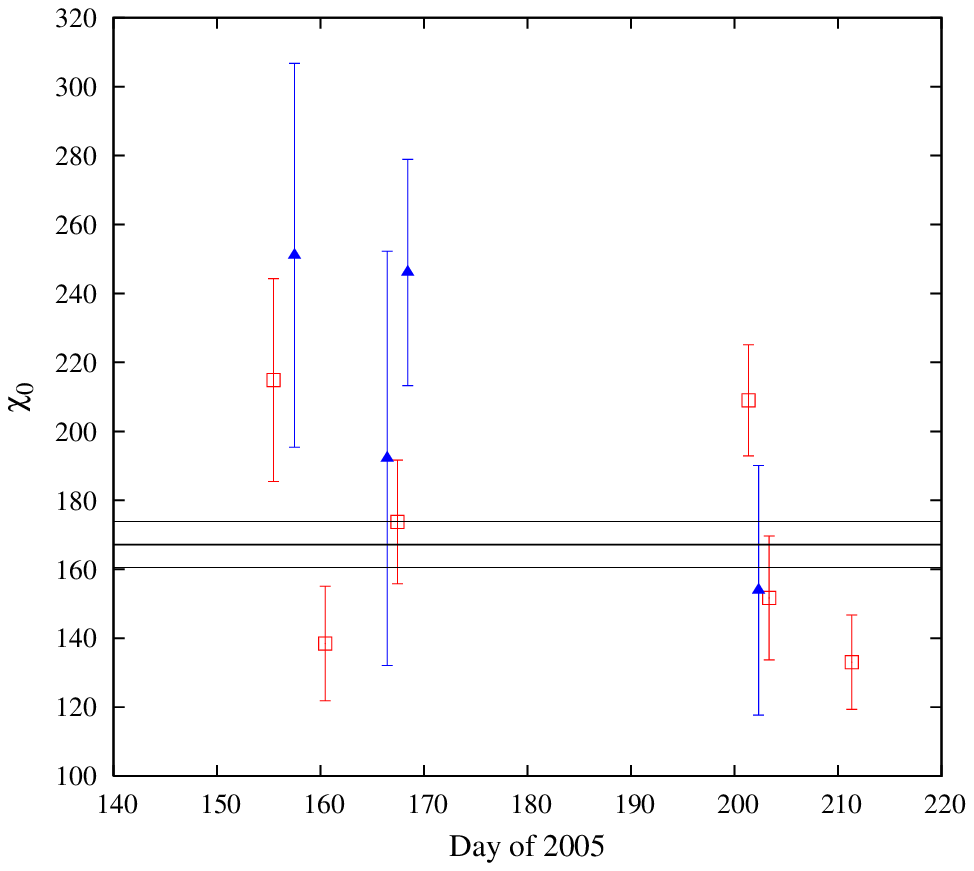}
\caption{\label{f-X0}Measurements of the intrinsic polarization
position angle ($\chi_0$) from the summer of 2005. The mean value is
$\chi_0=167^\circ\pm7^\circ$. Symbols as in
Figure~\ref{f-RM}.\vspace{2pc}}
\end{minipage} 
\end{figure}

Faraday rotation changes the observed polarization position angle
($\chi$) as a function of frequency according to:
\begin{equation}
\chi(\nu) = \chi_0 + \frac{c^2}{\nu^2}\mathrm{RM} ,
\label{e-RM}
\end{equation}
where $\chi_0$ is the intrinsic position angle. The rotation measure
(RM) is proportional to the integral of the electron density and
parallel magnetic field component along the line of sight. From the
observed LSB and USB position angles we derive a RM and $\chi_0$ for
each observation (see Figures~\ref{f-RM} and \ref{f-X0}). The larger
errors on 345~GHz points in these figures result from the much smaller
inter-sideband difference in $\lambda^2$ at the higher frequency. The
average RM from all 10 epochs is $(-5.6\pm0.7)\times10^5$~\rmm, while
the 230~GHz and 345~GHz points alone yield
$(-5.4\pm0.7)\times10^5$~\rmm\ and $(-13\pm5)\times10^5$~\rmm,
respectively, consistent within their errors. The ten single-night RM
values are consistent with a constant value ($\chi^2_r = 1.09$, for 9
degrees of freedom). We therefore place an upper limit of
$7\times10^4$~\rmm\ on the RM dispersion. 

This represents the first statistically significant measurement of the
RM of \sgra, and the only measurement made from simultaneous
observations at multiple frequencies and therefore able to isolate
source polarization changes. Our ten measurements show an average
intrinsic polarization of $167^\circ\pm7^\circ$, or
$162^\circ\pm7^\circ$ and $210^\circ\pm21^\circ$ from 230 and 345~GHz
observations, respectively. The $\chi_0$ values vary by more than our
measurement errors predict, suggesting intrinsic polarization
changes. Assuming a constant $\chi_0$, we obtain $\chi^2_r=2.8$ (0.3\%
probability) using all data points, or $\chi^2_r=3.9$ (0.16\%) for the
230~GHz points only. The variability in the full data set suggests an
intrinsic $\chi_0$ dispersion of $31^{+18}_{-9}$ degrees, with very
similar results obtained from 230 or 345~GHz points separately.

\subsection{Implications for the Nature of \sgra}
The RM we observe is too large to be produced by material beyond the
Bondi accretion radius of \sgra\ (approximately 1\arcsec\ or
0.04~pc). Using the density measurements of Baganoff et
al. \cite{BaganoffE03}, the RM in the inner 10\arcsec\ is just
$8\times10^3$~\rmm\ assuming a 1~mG ambient field. Observations of
nearby radio sources have found RMs up to 70\% of this estimate
\cite{YZMorris87}.

\begin{figure}[t]
\includegraphics[width=20pc]{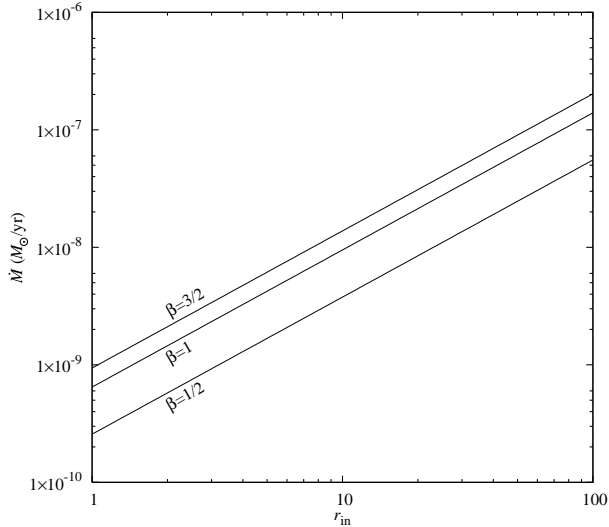}
\hspace{1.5pc}
\begin{minipage}[b]{16pc}
\caption{\label{f-mdot}Accretion rate constraints from the RM
detection, based on the Marrone et al. \cite{MarroneE06} formalism. We
plot the accretion rate as a function of \rin, the radius at which
electrons become relativistic, as this parameter strongly influences
the derived limits. The three lines trace three density profiles
bounded by the ADAF ($\beta=3/2$) and CDAF (1/2) models. We asssume a
coherent magnetic field ($r_\mathrm{out}=\infty$). If we instead
choose $r_\mathrm{out}=3r_\mathrm{in}$, all three lines are within a
factor of 2 of the $\beta=3/2$ line in this figure.}
\end{minipage}
\end{figure}

The RM can be used to determine the accretion rate at small radii
around \sgra\ if assumptions are made about the nature of the
accretion flow. The procedure is outlined in Marrone et
al. \cite{MarroneE06} and assumes a power-law radial density profile
($n\propto r^{-\beta}$) and ordered, radial, equipartition-strength
magnetic fields. The accretion rates allowed by our RM detection are
shown in Figure~\ref{f-mdot}. The most important parameter in
determining the range of accretion rate upper limits is \rin, the
radius at which electrons become relativistic and their RM
contribution begins to be suppressed \cite{QuatGruz00-LP}. In general,
shallower density profiles ($\beta\rightarrow 1/2$) approaching that
of the convection-dominated accretion flow (CDAF)
\cite{QuatGruz00-CDAF} yield hotter central temperatures and thus
larger \rin\ \cite{Quataert03}. Simulations favor $\beta$ values
closer to this type of flow (i.e., $\beta\leq1$)
\cite{YuanE03,HawleyBalbus02,IgumenshchevE03,PenE03}. Simulations with
published temperature profiles show \rin\ between 30 and
$100r_\mathrm{S}$ \cite{YuanE03,HawleyBalbus02}, yielding accretion
rate upper limits of $2\times10^{-7}$ to
$5\times10^{-8}$~\msunyr. Steeper profiles ($\beta\rightarrow3/2$), as
in Bondi-like or advection-dominated accretion flows (ADAF)
\cite{NaraYi95} are marginally relativistic to small radii and may
have \rin\ of a few $r_\mathrm{S}$, indicating an accretion rate upper
limit of $1.5\times10^{-8}$ for $r_\mathrm{in}=10r_\mathrm{S}$. This
has been used to rule out the original ADAF model, which has an
accretion rate comparable to the Bondi rate ($10^{-5}$~\msunyr\ for
this source, \cite{BaganoffE03}).

It is very important to note that the assumptions of the Marrone et
al. \cite{MarroneE06} formalism, which were taken from previous
related works \cite{QuatGruz00-LP,Melia92}, are not observationally
constrained. In particular, the assumption of equipartition-strength
fields is not well justified, although simulations may show a tendency
to develop field strengths of a few percent of equipartition
\cite{IgumenshchevE03}. Magnetic fields that are a fraction
$\epsilon$ of the equipartition strength will raise the accretion rate
limits by $\epsilon^{2/3}$ (a factor of 10 for $\epsilon=3$\%). The
assumption of an ordered field is also potentially suspect. The
direction of position angle change with frequency, and thus the sign
of the RM, appears unchanged since the Aitken et al. \cite{AitkenE00}
measurements in 1999. This sign stability in the RM and stability of
the sense of circular polarization may be evidence of a highly
reversed field with a small bias superimposed in an accretion flow or
jet \cite{RuszBegel02,BeckertFalcke02}. Sign stability would not be
expected from turbulent accretion flows unless a special geometry is
invoked or a rigid field is assumed \cite{IgumenshchevE03}. If the
stochastic field picture is correct, the accretion rate could be much
higher than the limits we have derived.

Our RM also allows us to place lower limits on the accretion rate;
these are not subject to the caveats above since all of the
uncertainties act to raise the minimum accretion rate. If we take
\rin\ to be around $10r_\mathrm{S}$ or $3r_\mathrm{S}$ (smaller \rin\
yields smaller lower limits, so this is conservative for hot flows),
we find that the accretion rate must be greater than
$1-2\times10^{-8}$~\msunyr\ or $2-4\times10^{-9}$~\msunyr,
respectively. This rules out a thin disk in this source, as accretion
rates of $10^{-9}$~\msunyr\ or less are required to match the
luminosity of \sgra\ with a thin disk \cite{Narayan02}. Moreover, if
the field is toroidal, reversed, or sub-equipartition, these lower
limits are raised and may pose problems for very low accretion rate
models. Our limit on RM variability (12\%) also limits the fractional
accretion rate variability to 8\% over two months.

\section{Conclusions}
We have shown several new observations of \sgra, made possible by the
Submillimeter Array and its polarimeter. In particular, we report the
first significant detection of the RM in this source and discuss its
implications. Significant work remains in polarimetric studies of
\sgra. In particular, these observations cannot be used to detect
deviations from the $\nu^{-2}$ dependence expected for Faraday
rotation. Such deviations, arising from internal Faraday rotation,
would be evidence for cooler than expected gas at small radii, which
would favor steep density profiles. Future dual-receiver SMA
observations or joint SMA-CARMA polarimetry could yield measurements
at three or more frequencies and resolve this question.

\ack
We thank the Submillimeter Array staff, particularly Ken Young, for
vital assistance in developing the polarimetry system. We also thank
the organizing committee for a well-managed and useful conference.

\end{document}